# Magnetism in Transition metal doped Cubic SiC

Padmaja Patnaik, Gautam Mukhopadhyay and Prabhakar P. Singh

*Physics Department, I.I.T.Bombay, Powai, Mumbai 400076, India.*
*Email: padmaja@phy.iitb.ac.in*

**Abstract.** We report here our study on SiC doped with transition metals using first principle density functional theory calculations. We have considered cubic SiC with 3d transition metals as substitutional impurities for Si and C site separately. Cubic SiC doped with Cr, Mn, show ferromagnetism whereas with Sc, Ti, V and Co show site dependency of magnetic properties. Rests of the impurities are found to be non-magnetic.

## INTRODUCTION

The properties of a semiconductor changes, in fact gets enhanced with the presence of certain impurities. When a small concentration of magnetic material, in general transition metal (TM) atom is added to a semiconductor compound, it behaves as a Dilute Magnetic Semiconductor (DMS) [1]. The DMS have found their application in Spintronics devices [2] replacing the usual magnetic materials. The indirect band gap semiconductor SiC has found its applications in high power devices because of its properties like high thermal conductivity, high hardness and high saturation value of electron drift velocity. The TM when added to SiC gives rise to electrically active centers in it. TM can be found in SiC as impurities during growth [3] and also can be added intentionally [4] to develop a DMS of SiC. The SiC DMS has found its applications in high power and high temperature electronics. Here we have considered a number of TMs doped in cubic SiC as substitutional impurities in Si or C place. We discuss the magnetic properties of cubic SiC doped with a number of TMs by using first principles, density functional theory (DFT).

## METHODOLOGY

All calculations are done with a first principle energy code, Quantum Espresso [5] which uses pseudopotentials within density functional theory. This code performs the density functional pseudopotential total energy calculations in periodic systems solving iteratively the Kohn-Sham equations in a plane wave basis set. Both norm-conserving and Vanderbilt ultra-soft pseudopotentials are used. Calculations are based on DFT in the generalized gradient approximation (GGA) suggested by Perdew, Burke and Ernzerhof (PBE) [6]. We considered one of the most popular polytype cubic SiC. We have used super cells of different sizes for impurity calculations to minimize the impurity interactions. Here we present the results obtained from a super cell obtained by expanding the unit cell three times in all directions. The expansion of wave function in a plane wave basis set is restricted by kinetic energy cut-off of 28 Ry. The value of cut-off energy and size of k-point mesh is obtained by total energy convergence method. The calculated value of lattice constant is 4.334Å but we have used the experimental value 4.358Å [7].

## RESULTS AND DISCUSSIONS

We have considered cubic SiC as the host and few TMs as substitutional impurity in Si and C place separately. We have done spin polarized calculations in all the cases to check the magnetic behavior of the sample. The distance between two consecutive impurity atoms is 13.08Å.

### Magnetic Behavior

To understand the magnetic behavior of the sample we have looked for the magnetic moment of the TM in the doped SiC and also the total magnetic moment of the sample after doping. As seen in Fig. 1 the magnetic moment of TM behaves differently in Si and C sites. For early TMs in the series, the C site substitution show higher magnetic moments than the Si site substitution. But as one moves forward in the series, heavy TMs, after Fe, show paramagnetic solution for C site. Situation is different in Si site substitution as all elements after Fe show positive magnetic moment values except Cu and Zn. Cr and Mn shows high magnetic moment values for both site substitutions. The first element in the TM series Sc shows totally site dependent magnetic behavior with highest magnetic moment in C site case and almost null magnetic moment in Si site case [8, 9]. This is true for three initial TMs, Sc, Ti and V [10].

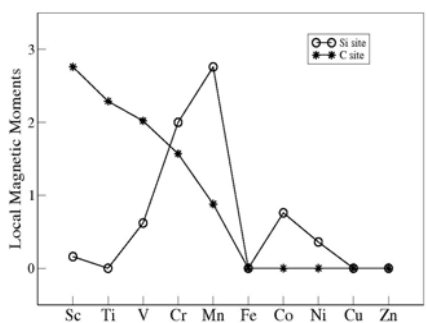

**FIGURE 1**. Magnetic moment of the impurity atoms are plotted in Bohr magnetron units.

To understand the spin state of these impurities in SiC we can have a look at the density of states (DOS). Fig. 2 shows the DOS and partial DOS of few of the impurities. Impurity levels inside the band gap due to d-electrons of the impurity atom. Both Cr and Mn shows high spin state. Sc, Ti and V create impurity levels inside band gap for C site substitution [11, 12] and Co creates impurity level for Si site substitution.

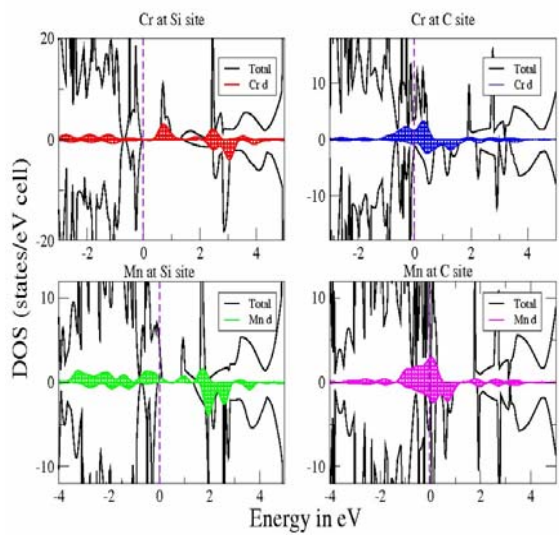

**FIGURE 2.** Total DOS and PDOS of TMs at different sites in cubic SiC are shown. Valence band maximum is taken at zero of energy axes.

## CONCLUSIONS

We find that TM impurity level in cubic SiC depend on the substitutional site. Cr, Mn and V are found to form ferromagnetic solution with cubic SiC and hence can be probable candidates for DMS. Sc and Ti shows high magnetic moment when replaces a C atom but not in case of Si replacement. Hence magnetic properties introduced in cubic SiC not only depend on the added impurity atom but also on the site of substitution.